\documentclass[conference]{IEEEtran}

\usepackage{amsmath,amssymb,amsfonts}
\usepackage{algorithmic}
\usepackage{graphicx}
\usepackage{textcomp}
\usepackage{xcolor}
\usepackage{orcidlink}

\usepackage[utf8]{inputenc}
\usepackage[T1]{fontenc}
\usepackage{cleveref}
\usepackage{url}

\usepackage{breakurl}
\usepackage{hyperref}
\hypersetup{hidelinks,unicode,breaklinks}
\usepackage[style=ieee, backend=biber, url=false, hyperref, maxnames=5, minnames=1, maxcitenames=2, mincitenames=1, defernumbers=true]{biblatex}
\usepackage{acronym}
\usepackage{dblfloatfix}
\usepackage{tabto}
\usepackage{enumerate}
\usepackage{siunitx}

\usepackage{tikz}
\usetikzlibrary{math,arrows.meta,calc} 

\acrodef{AES}{Advanced Encryption Standard}
\acrodef{ECF}{Encrypted Container File}
\acrodef{GCM}{Galois/Counter Mode}
\acrodef{GPG}{GNU Privacy Guard}
\acrodef{KDF}{Key Derivation Function}
\acrodef{PoC}{Proof of Concept}

\newcommand{\gc}{\textit{git-crypt}}
\newcommand{\jak}{\textit{jak}}
\newcommand{\cs}{\textit{C\#}}
\newcommand{\at}{\makeatletter@\makeatother}
\newcommand{\ecf}{\ac{ECF}}
\newcommand{\arr}[1]{[#1]}
\newcommand{\ba}[1]{Byte Array~\arr{#1}}

\newcommand{\func}[1]{\text{#1}}
\newcommand{\f}[3]{\func{#1}_{\textnormal{#2}}^{\textnormal{#3}}}
\newcommand{\var}[3]{#1_{\textnormal{#2}}^{\textnormal{#3}}}
\newcommand{\sk}[2]{\f{sk}{#1}{#2}}
\newcommand{\pk}[2]{\f{pk}{#1}{#2}}
\newcommand{\aes}[1]{\f{k}{#1}{AES}}
\newcommand{\trunc}[2]{[#1,\!...,\!#2]}

\newcommand{\sss}[1]{\vspace{2mm}\subsubsection{#1}~\vspace{0mm}\\}

\newenvironment{fielditemize}{\itemize \setlength{\itemsep}{0.25em}}{\enditemize}

\crefname{subsection}{subsection}{subsections}
\Crefname{subsection}{Subsection}{Subsections}

\usepackage[style=ieee, backend=biber, url=false, hyperref, maxnames=5, minnames=1, maxcitenames=2, mincitenames=1, defernumbers=true]{biblatex}

\makeatletter
\renewcommand{\fnum@figure}{Figure~\thefigure}
\makeatother

\usepackage{ifpdf}
\ifpdf
\pdfoutput=1 
\pdftrue
\fi

\makeatletter
\def\ps@IEEEtitlepagestyle{
  \def\@oddfoot{\mycopyrightnotice}
  \def\@evenfoot{}
}
\def\mycopyrightnotice{
  {\footnotesize
    \begin{minipage}{0.8\textwidth}
    \centering
    Please cite as: \fullcite{selfref}.
    \end{minipage}
  }
}
\makeatother
\makeatletter
\let\blx@rerun@biber\relax
\makeatother
\usepackage[shortcuts]{extdash} 

\DeclareBibliographyCategory{selfref}
\addtocategory{selfref}{selfref}

\title{\bfseries\Large Encrypted Container File:\\
Design and Implementation of a Hybrid-Encrypted Multi-Recipient File Structure
}

\author{%
    \large Tobias J. Bauer\,\orcidlink{0009-0006-1073-3971} and Andreas Aßmuth\,\orcidlink{0009-0002-2081-2455}\\[0.3ex]\normalsize\normalfont
	Ostbayerische Technische Hochschule Amberg-Weiden\\
	Department of Electrical Engineering, Media and Computer Science\\
	Kaiser-Wilhelm-Ring 23, 92224 Amberg, Germany\\
	Email: {\tt \{t.bauer$\,|\,$a.assmuth\}@oth-aw.de}%
}

\begin{document}

\maketitle

\begin{abstract}
Modern software engineering trends towards Cloud-native software development by international teams of developers. Cloud-based version management services, such as GitHub, are used for the source code and other artifacts created during the development process. However, using such a service usually means that every developer has access to all data stored on the platform. Particularly, if the developers belong to different companies or organizations, it would be desirable for sensitive files to be encrypted in such a way that these can only be decrypted again by a group of previously defined people.
In this paper, we examine currently available tools that address this problem, but which have certain shortcomings. We then present our own solution, Encrypted Container Files (ECF), for this problem, eliminating the deficiencies found in the other tools.
\end{abstract}

\renewcommand\IEEEkeywordsname{Keywords}

\section{Introduction}\label{sec:introduction}

Software development undergoes a permanent change and, occasionally, long-lasting trends emerge, which influence the choices made in terms of software architectures, technologies, programming languages and frameworks used. Current trends involve the development of Cloud-native distributed software components which are deployed automatically via Continuous Delivery and Continuous Deployment~\autocite{Intellectsoft2021}.

This implies that these components, often running in separate containers, must communicate with each other. Furthermore, there is an interest in securing such communication links because very often confidential data is transmitted. This in turn places demands on the software development process: in order to secure (digital) communications these must be encrypted. This is also true for storing confidential data. In common cases, e.g., running a web server or storing confidential data in a database, means of authentication must be kept secret. Such means of authentication include, but are not limited to, passwords, private certificate keys, and symmetric encryption keys.

Modern software development takes place in teams whose members are in constant exchange with each other. Often, version control systems, e.g., \textit{git}~\autocite{Git2022} are used to manage the source code and other artifacts. Also with regard to the practice of Continuous Integration (see~\autocite{Fowler2006}), which is a preliminary step to the aforementioned Continuous Delivery and Continuous Deployment, it is necessary to check-in \textit{all} artifacts into the version control system. This would be grossly negligent for confidential data provided that no protective measures against unauthorized access are taken.

In this paper, we address the issue of access to an encrypted file structure in the cloud by different people in a software development team. With the \ecf, we present our own solution for a cloud-based, encrypted data storage for software development teams, in which the functionality of currently available tools is extended and their shortcomings are eliminated.

This paper is structured as follows: in \Cref{sec:relatedwork}, there is a brief introduction to two existing solutions before the requirements are presented in \Cref{sec:requirements}. In \Cref{sec:struct}, we present an example of use and describe the structure and operations of the \ecf. Following that, \Cref{sec:implementation} describes implementation details. Finally, \Cref{sec:conclusion} concludes the paper and gives an outlook on future work.

\section{Related Work}\label{sec:relatedwork}

There are different solutions to address the issue we described in \Cref{sec:introduction}. In this section, we give an overview of two of these tools, \jak\ and \gc, and discuss their features and shortcomings.


The tool \jak~\autocite{Dispel2017} is written in Python and allows symmetric encryption of files using \ac{AES}. Using the tool, one can generate keys and store them in a keyfile, which is not encrypted. To enable automatic encryption and decryption with a single command \jak\ uses a special text file that contains a list of file names. This special text file can be added to the repository~\autocite{Dispel2017}.

The practical use is limited because of sole symmetric encryption as the key distribution problem remains unsolved. Especially with growing team sizes distributing confidential data results in disproportionate effort.

Another issue with \jak\ is that the confidential files' content stays unencrypted on the developers' computers. This is because \jak\ decrypts these files during checkout and re-encrypts them before committing. This implies that only externals with reading access to the repository at maximum and no access to any of the developers' computers are unable to access the confidential data. A common application scenario are projects that are developed on a public repository platform.


The tool \gc~\autocite{Ayer2022} allows symmetric encryption of files within a \textit{git} repository using \ac{AES}, too. It shares the same limitations as \jak\ in terms of access restrictions to externals.

However, \gc\ offers a solution to the key distribution problem by using \ac{GPG}~\autocite{Koch2022}. Public \ac{GPG}-keys, the recipients' keys, can be added to the repository. When encrypting the confidential files within the repository \gc\ generates an asymmetrically encrypted keyfile for each recipient. Every recipient therefore gets access to the symmetric key and because of that is capable of decrypting the confidential files in that repository.

The tool \gc\ is implemented in a way that all confidential files are encrypted with the same symmetric key and this very key must therefore be shared with all recipients added to the repository. This results in coarse grained access control as there is no way to restrict access to some confidential files to a subset of the recipients. 
Consider the case that, e.g., secret information about the production environment should only be accessible to the production team.

Furthermore, \gc\ does not secure the confidential files' content on the developers' computers. This is analogous to \jak\ because both tools decrypt the confidential files during checkout. The integration of \gc\ into the mechanisms of \textit{git} is optional but recommended~\autocite{Ayer2022}.

Another shortcoming of \gc\ is the lacking feature to remove recipients. \citeauthor*{Ayer2022} justifies this by stating that by using a version control system a removed recipient can still access old versions of the repository and, therefore, the confidential data stored within~\autocite{Ayer2022}. This argument is correct as far as it goes~-- nevertheless, it seems sensible to implement such a mechanism into the to-be-designed \ecf\ format since the confidential data should be updated regularly regardless. For example, certificates and passwords expire and symmetric keys should be changed regularly with regards to staff turnover.

\section{Requirements Engineering}\label{sec:requirements}

From the features and shortcomings of the \jak\ and \gc\ tools presented in \Cref{sec:relatedwork}, some requirements for the \ecf\ format can be derived:

\begin{enumerate}
    \item\label{e:req:enc} Mandatory encryption of confidential data,
    \item\label{e:req:mod} possibility to modify confidential data (content is writable),
    \item\label{e:req:key} key distribution is no prerequisite,
    \item\label{e:req:dec} decryption not during checkout but on demand,
    \item\label{e:req:mul} support for multiple recipients,
    \item\label{e:req:arr} addition and removal of recipients,
    \item\label{e:req:mig} minimal information gain for external parties, and
    \item\label{e:req:set} customizable set of recipients per file.
\end{enumerate}

Based on these requirements, we have decided to use the following design goals for our solution:

\begin{itemize}
    \item Use of hybrid encryption (\Cref{e:req:enc,,e:req:key,,e:req:mul}),
    \item inclusion of recipient information to allow re-encryption on changes (\Cref{e:req:mod,,e:req:mul,,e:req:arr,,e:req:set}),
    \item obfuscation of recipient information for respective external parties (\Cref{e:req:mig,,e:req:set}), and
    \item delivery of the associated software as a library for embedding into existing applications (\Cref{e:req:dec}).
\end{itemize}

\begin{figure*}[!hb]
\vspace*{-5mm}
\centering
\begin{tikzpicture}
    \tikzset{arr/.style={color=border, -latex}}
    \input{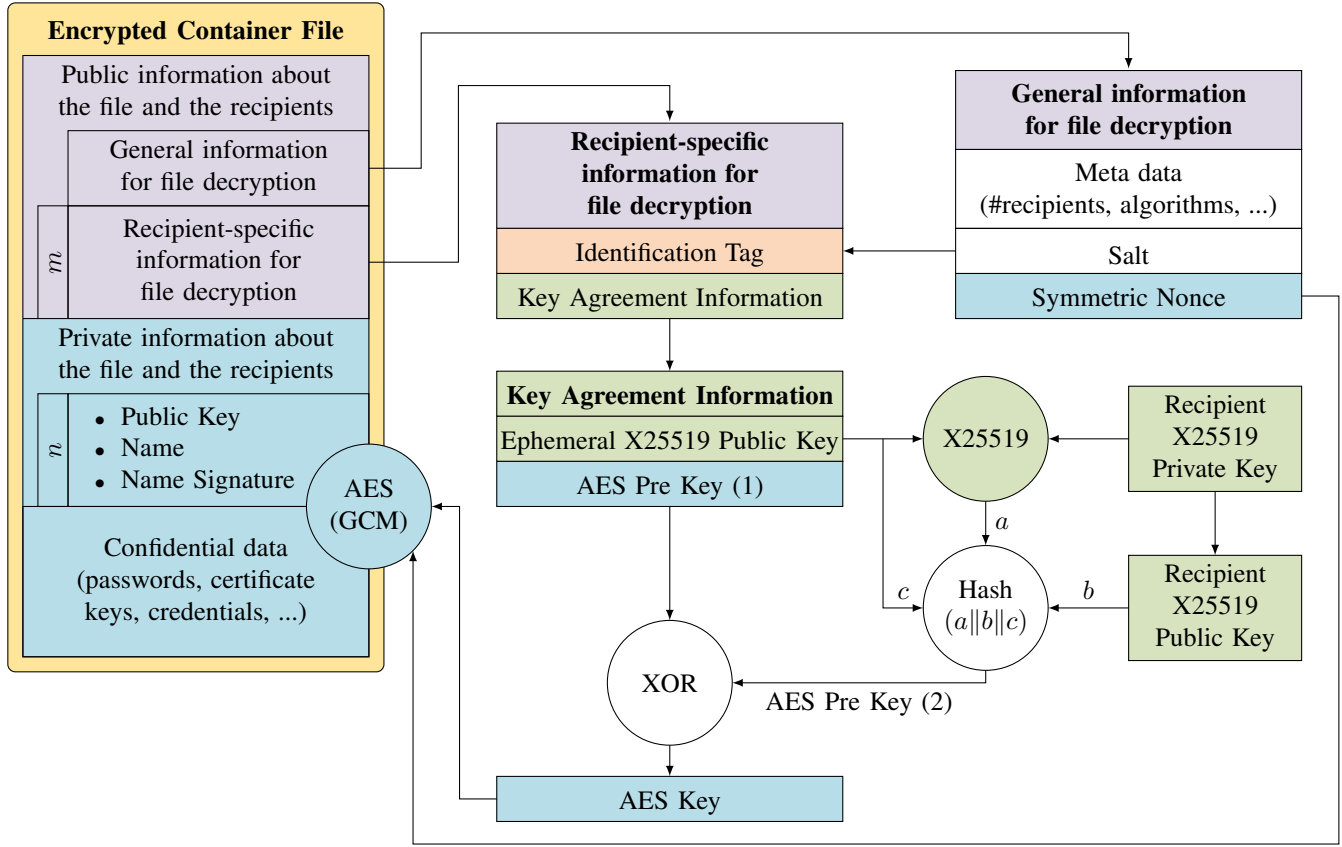}
\end{tikzpicture}
\vspace*{-2mm}
\caption{Diagram of the most important components of an \acl{ECF} and visualization of the interrelationships.}\label{fig:components}
\vspace*{-4mm}
\end{figure*}

\section{Structure and Operations of the Encrypted Container File}\label{sec:struct}

This section gives an overview over the use of the \ecf\ format in \Cref{sec:struct:usage}. The following subsections describe the structure of the \ecf\ format in detail. \Cref{fig:components} shows an overview of the components of an \ecf, how they are connected and related to each other. \Cref{sec:struct:general} describes the general structure, components, and storage formats of an \ecf. The publicly accessible fields are described in \Cref{sec:struct:public} and the private fields in \Cref{sec:struct:private}. The following \Cref{sec:struct:decryption,sec:struct:encryption} describe the decryption and encryption process, respectively. Finally, \Cref{sec:struct:operations} concludes this section with further operations that can be performed on an \ecf.

\subsection{Usage in Practice}\label{sec:struct:usage}

In this subsection, we walk through the following scenario: Alice wants to encrypt a file using the \ecf\ format and operations in such way that her friend Bob will be able to read the content, while Charlie should not be able to.

First, Alice needs access to Bob's public information, which comprises among others his public key. Bob must have created his public information beforehand. Next, Alice creates an \ecf\ using, e.g., the CLI tool described in this paper and provided via GitHub, and adds the confidential data. After that, she can add Bob as a recipient using his public information. To retain access to the content, Alice should add herself as a recipient to the \ecf. Alice can now save the \ecf\ within a public repository and only Bob and herself are able to decrypt the file's content. Charlie, on the other hand, cannot retrieve the encryption key as he is not a recipient of that \ecf\ and has therefore no access to the confidential data stored inside.

\subsection{General Structure and Data Type Storage Format}\label{sec:struct:general}

Each \ecf\ consists of three parts: A public part and two non-public/private parts. In \Cref{fig:components}, the whole \ecf\ is framed yellow, whereas the public part is colored purple. Both private parts are treated as a single datum by the symmetric encryption and are colored in blue. The following list describes the data types used in the following subsections and their storage format within an \ecf:

\begin{itemize}
    \item \texttt{\small Unsigned Integer}: \tabto{3.6cm} 4 Bytes, Little Endian
    \item \texttt{\small \ba{x}}:           \tabto{3.6cm} $x$ Bytes, sequential
    \item \texttt{\small String}:           \tabto{3.6cm} 4 Bytes, little endian (Length),\\
                                    \tabto{3.6cm} then UTF-8 bytes without \\
                                    \tabto{3.6cm} byte order mark (BOM)
\end{itemize}

The \ecf\ format is designed to be flexible with regards to the used cipher suite. In order to allow future extensions, it allows more algorithms and cipher suites. For this paper and also for our \ac{PoC} implementation, a selection for the cipher suite was made, which is the basis for the rest of this paper:

\begin{itemize}
    \item Key Agreement/Exchange: \tabto{4cm} X25519~\autocite{Bernstein2006}
    \item Symmetric Encryption:   \tabto{4cm} \ac{AES}-256-GCM~\autocite{McGrew2005}
    \item Signature:              \tabto{4cm} Ed25519~\autocite{Bernstein2012}
    \item Hash Function:          \tabto{4cm} SHA-512~\autocite{Dang2015}
\end{itemize}

\subsection{Public Fields}\label{sec:struct:public}

Each \ecf\ must provide enough information for all authorized recipients to decrypt the file. Information for encrypting, however, is not required to be public because only recipients should be able to modify the confidential data within the \ecf. Hence, the public part of an \ecf\ contains just the information required for decryption. It comprises a general part and then $m$ identically constructed recipient-specific parts.

The general part contains the following data (in this order):

\begin{fielditemize}
    \item \texttt{\small Container Version (Unsigned Integer)}\\\ecf\ format version; intended for future extensions
    \item \texttt{\small Cipher Suite (Unsigned Integer)}\\Information about used algorithms
    \item \texttt{\small Public~Header~Length~(Unsigned~Integer)}\\Length of the public part in Bytes
    \item \texttt{\small Private Length (Unsigned Integer)}\\Length of the private part in Bytes
    \item \texttt{\small Recipient Count (Unsigned Integer)}\\Number of recipients in the public part ($m$)
    \item \texttt{\small Salt (\ba{16})}\\Salt value (usage described below)
    \item \texttt{\small Symmetric Nonce (\ba{12})}\\Symmetric nonce value (usage described below)
\end{fielditemize}

The first two fields, \texttt{\small Container Version} and \texttt{\small Cipher Suite}, are used to make the \ecf\ format flexible and future-proof. However, we discuss only the cipher suite selected in \Cref{sec:struct:general}.

The recipient-specific decryption information is yet to be defined. In total, $m$ such blocks~-- one for each recipient~-- are stored after the general part. To obfuscate the number of recipients towards externals, $m \geq n$ can be chosen freely with $n$ the true number of recipients. Random blocks, which belong to no recipient, may be inserted, which is not evident to externals. Each recipient-specific block consists of two fields: an \texttt{\small Identification Tag (\ba{16})}, which is used to assign a block to a recipient, and \texttt{\small Key Agreement Information} that contains recipient-specific decryption information.

The field \texttt{\small Identification Tag} is colored orange in \Cref{fig:components}. It is the hash value truncated after 16 Bytes from the concatenation of the bit strings of the public key of the respective recipient and the \texttt{\small Salt} value introduced above. Shortening the hash value saves storage space and allows with $(2^8)^{16}=2^{128}$ possible values for practically unlimited unique recipients. An authorized recipient can calculate their \texttt{\small Identification Tag} based on the knowledge of their own public key and the public \texttt{\small Salt}.


The second field, \texttt{\small Key Agreement Information}, contains recipient-specific information for the decryption process and is highly dependent on the used cipher suite. For the selected cipher suite, an \texttt{\small Ephemeral X25519 Public Key (\ba{32})} and an \texttt{\small \ac{AES} Pre Key (\ba{32})} is stored. The first is used in the key agreement phase to obtain a second \ac{AES} pre key, the latter is the first \ac{AES} pre key. \Cref{sec:struct:decryption} describes the combination of the available recipient-specific information to obtain the symmetric key in more detail. The process is depicted in \Cref{fig:components} as well.

\subsection{Private Fields}\label{sec:struct:private}

The private part of an \ecf\ consists of two segments: First, there is information about the \ecf\ and its recipients, and second, there is the encrypted confidential data itself. The private part is completely encrypted symmetrically and, therefore, not accessible for external parties. The following fields are stored in the private part:

\begin{fielditemize}
    \item \texttt{\small Content Type (Unsigned Integer)}\\Describes the type of the confidential data
    \item \texttt{\small Public Header Hash (\ba{64})}\\Hash value of the public part
    \item \texttt{\small Recipient Count (Unsigned Integer)}\\Number of true recipients ($n$)
    \item \texttt{\small Recipient Information (Array \arr{n})}\\Information about the recipients ($n$ blocks)
    \item \texttt{\small Content Length (Unsigned Integer)}\\Length of the confidential data in Bytes ($len$)
    \item \texttt{\small Content (\ba{len})}\\Confidential data
    \item \texttt{\small Private Hash (\ba{64})}\\Hash value of the private part so far
\end{fielditemize}

The first field, \texttt{\small Content Type}, is intended for future use and should characterize the type of confidential data stored in \texttt{\small Content}. It seems reasonable that future applications using the \ecf\ library will define and handle their own content types.

The field \texttt{\small Public Header Hash} contains the hash value over the whole public part of an \ecf. The value for the public field \texttt{\small Public Header Length} (c.f.~\Cref{sec:struct:public}) is unknown at the time of encryption because the length of the symmetric encryption algorithm's output is not necessarily known in advance. Therefore, this field is set to the constant value of \texttt{\small 0xECFFC0DE} (\acl{ECF} Format Code) during the calculation of the hash value. The \texttt{\small Public Header Hash} is used to detect unauthorized or unintended modifications of the public part, e.g., non-destructive changes of recipient-specific information of other recipients.

The fields \texttt{\small Recipient Count} and \texttt{\small Content Length} specify the number of true recipients and the length of the confidential data, respectively. The field \texttt{\small Recipient Information} consists of $n$~blocks of variable length, which in turn consist of three fields: the \texttt{\small Public Key (\ba{32})} of the recipient, a \texttt{\small Name (String)} which contains a self-chosen name of the recipient (variable length), and the \texttt{\small Name Signature (\ba{64})} over the self-chosen name.

Every block of \texttt{\small Recipient Information} contains information about a recipient, so that re-encrypting the \ecf\ is possible, e.g., after modifying confidential data. These information blocks about the recipients are stored within the private part of the \ecf\ in order to hide them from externals. The block field \texttt{\small Public Key} contains the recipient's public key, which is a public Ed25519 key as specified in \Cref{sec:struct:general}. One can convert an Ed25519 public key into an X25519 public key as described in \autocite{CSE2019}\autocite{Sodium2022ed2c}. The next block field \texttt{\small Name} holds a text of variable length that describes the recipient. It may contain the name of the related person or their email address. This field is for human legibility and information purposes only, e.g., when displaying the recipients or when removing existing recipients. The last block field, \texttt{\small Name Signature}, contains a signature over the content of \texttt{\small Name}. The signature is used first and foremost to ensure, that the person owing the associated private key has chosen the name, and that no changes have been made to the name by other recipients afterwards.

The field \texttt{\small Content} encloses the confidential data and has a theoretical limit of $2^{32} - 1 \mathop{}\textnormal{Bytes} \approx 4\mathop{}\textnormal{GiB}$. In practice, this limit should never be reached because an \ecf\ is designed primarily to be used with passwords, certificate keys, credentials and similar confidential data.

The last field, \texttt{\small Private Hash}, takes the hash value over the private part up to this point. This field is inside the private part of an \ecf\ and, therefore, the hash value is calculated before encryption. A more detailed description of the encryption process can be found in \Cref{sec:struct:encryption}.

\subsection{Decryption Process}\label{sec:struct:decryption}
This subsection describes the processes of calculating the \ac{AES} key according to \Cref{fig:components}. To decrypt an \ecf\ a recipient needs both, their private X25519 key and their public Ed25519 key. Both can be calculated from the recipient's private Ed25519 key~\autocite{CSE2019}\autocite{Sodium2022ed2c}.

\textit{Nomenclature.} The following notation is used: \textit{Alice} is the recipient and $\sk{A}{Ed}$ denotes her private Ed25519 key, $\pk{A}{X}$ denotes her public X25519 key, analogously. The used cryptographic hash function is denoted by $\func{H}$, $a \Vert b$ denotes the concatenation of two bit strings $a$ and $b$, and $a \oplus b$ denotes the bitwise exclusive OR (XOR) operation on two bit strings $a$ and $b$ of the same length. $a\trunc{0}{n}$ denotes the truncation of the bit string $a$ to the first $n$ Bytes. The ephemeral public X25519 key contained in the recipient-specific decryption information is denoted by $\pk{e}{X}$. The function $\func{X25519}(a, B)$ describes the multiplication of scalar $a$ with point $B$ on the elliptic curve \textit{Curve25519}~\autocite{Bernstein2006}.

Alice performs the following steps to obtain the \ac{AES} key:

\Crefname{enumi}{Step}{Steps}
\begin{enumerate}[(1)]
    \setlength{\itemsep}{0.25em}
    \item\label{e:dec:id} Compute $\func{identification\_tag} = \func{H}\!\left(\pk{A}{Ed} \Vert \func{Salt}\right)\!\trunc{0}{16}$.
    \item\label{e:dec:load} Load the decryption information $\left(\pk{e}{X}, \aes{pre1}\right)$ with matching $\func{identification\_tag}$.
    \item\label{e:dec:ka} Execute the key agreement algorithm with Alice's private X25519 key and the public ephemeral X25519 key:\\ $\f{k}{shared}{X} = \func{X25519}\!\left(\sk{A}{X}, \pk{e}{X}\right)$.
    \item\label{e:dec:p2} Compute $\aes{pre2} = \func{H}\!\left(\f{k}{shared}{X} \Vert \pk{A}{X} \Vert \pk{e}{X}\right)\!\trunc{0}{32}$.\\Shortening the hash value to $32$ Bytes is necessary because of the used symmetric encryption algorithm \ac{AES}-256
    \item\label{e:dec:key} Compute $\aes{} = \aes{pre1} \oplus \aes{pre2}$.
\end{enumerate}

In \Cref{e:dec:p2} the hash function gets evaluated on the concatenation of the shared key and both public keys to obtain the second \ac{AES} pre key. The reason for this is a recommendation in \autocite{Sodium2022psm} to not use the shared key $\f{k}{shared}{X}$ directly but to transform it with a hash function first. The question arises to why a simple hash function is used and not a \ac{KDF}. Primarily, the reason is to speed up the encryption process, because using a \ac{KDF} is resource-intensive and it must be computed $n$ times (separately for each of the $n$ recipients). This would result in a far slower encryption process for large $n$. Furthermore, the input data in \Cref{e:dec:p2} is substantially longer than the symmetric pre key to be computed, which makes key stretching not required and, therefore, the use of a cryptographic hash function seems sufficient.

Finally, the private part of an \ecf\ can be decrypted by using the computed \ac{AES} key $\aes{}$ and the public \ac{AES} nonce. In this paper, the \ac{GCM}~\autocite{Dworkin2007} was chosen for the symmetric encryption algorithm \ac{AES}. Therefore, one is not required to check the authenticity of the decrypted data separately. Furthermore, instead of the field \texttt{\small Public Header Hash} the public part of the \ecf\ could have been authenticated with \ac{AES}-\ac{GCM}. However, when supporting different modes of operation this field would have been required anyway. Hence, the field \texttt{\small Public Header Hash} was not removed and no additional data (\textit{Associated Data}) was added to the \ac{AES}-\ac{GCM} encryption algorithm.

\subsection{Encryption Process}\label{sec:struct:encryption}

The encryption process consists of an initial key and nonce generation step and an $m$-wise computation of the public X25519 ephemeral keys and \ac{AES} pre keys. For each of the $n \leq m$ true recipients exactly one public recipient-specific decryption information block must be generated. The remaining $m-n$ blocks serve as obfuscation and may be generated using a special process as proposed in \ref{sec:appendix:obfuscation}.

\textit{Nomenclature.} The same nomenclature applies as in \Cref{sec:struct:decryption}. It gets extended by the following functions: $\f{Gen}{}{AES}(256)$ and $\f{Gen}{}{X}$ denote functions to generate \ac{AES}-256 keys and X25519 key pairs, respectively. $\func{RandomBytes}(x)$ denotes a function to generate a random bit string of length $x$~Bytes.

For each recipient \textit{Bob}, their public Ed25519 key $\pk{B}{Ed}$ is known by every recipient of that \ecf\ because of the (private) block field \texttt{\small Public Key} (see \Cref{sec:struct:private}). Based on $\pk{B}{Ed}$ one can calculate Bob's public X25519 key $\pk{B}{X}$~\autocite{CSE2019}\autocite{Sodium2022ed2c}.

First, a symmetric \ac{AES} key $\aes{} \leftarrow \f{Gen}{}{AES}(256)$, an \ac{AES} nonce $\f{nonce}{}{AES} \leftarrow \func{RandomBytes}(12)$ and a bit string $\func{Salt} \leftarrow \func{RandomBytes}(16)$ must be generated at random (randomness indicated by the left arrow ``$\leftarrow$'').

Then, the following steps are performed $n$ times to generate the key agreement information for each recipient Bob:

\begin{enumerate}[(1)]
    \setlength{\itemsep}{0.25em}
    \item\label{e:enc:id} Compute $\func{identification\_tag} = \func{H}\!\left(\pk{B}{Ed} \Vert \func{Salt}\right)\!\trunc{0}{16}$.
    \item\label{e:enc:ge} Generate an ephemeral X25519 key pair:\\$\left(\sk{e}{X}, \pk{e}{X}\right) \leftarrow \f{Gen}{}{X}$.
    \item\label{e:enc:ka} Execute the key agreement algorithm with the private ephemeral X25519 key and Bob's public X25519 key:\\$\f{k}{shared}{X} = \func{X25519}\!\left(\sk{e}{X}, \pk{B}{X}\right)$.
    \item\label{e:enc:p2} Compute $\aes{pre2} = \func{H}\!\left(\f{k}{shared}{X} \Vert \pk{B}{X} \Vert \pk{e}{X}\right)\!\trunc{0}{32}$.\\Shortening the hash value to $32$ Bytes is necessary because of the used symmetric encryption algorithm \ac{AES}-256.
    \item\label{e:enc:p1} Compute $\aes{pre1} = \aes{} \oplus \aes{pre2}$.
\end{enumerate}

\Cref{e:enc:ge,,e:enc:ka} correspond to a ``half'' Diffie-Hellman key exchange~\autocite{DH1976} that gets completed during decryption (see \Cref{sec:struct:decryption}) in \Cref{e:dec:ka}.

For each recipient Bob the recipient-specific information can be written into the public part of the \ecf. This information per recipient consists of $\func{identification\_tag}$, public ephemeral X25519 key $\pk{e}{X}$ and \ac{AES} pre key $\aes{pre1}$.

The values $\func{Salt}$ and $\f{nonce}{}{AES}$ are valid for all recipients and are written into their respective fields (see \Cref{sec:struct:public}).

\subsection{Further \ecf\ Operations}\label{sec:struct:operations}

This subsection introduces more \ecf\ operations which are based on the elementary operations Decryption (\Cref{sec:struct:decryption}) and Encryption (\Cref{sec:struct:encryption}). The same nomenclature is used as in the specified subsections. It gets extended by the function $\f{Dec}{}{ECF}\!\left(\sk{A}{Ed}, \mathcal{E}\right)$ which denotes the decryption of an \ecf\ $\mathcal{E}$ with Alice's private Ed25519 key $\sk{A}{Ed}$. This function returns a tuple $(R, p)$ after successful decryption, with $R$ being the set of all $n$ recipients $R=\left\{r_1, r_2, \dots, r_n\right\}$ and $p$ being the bit string of the decrypted confidential data. Analogous to this, the function $\f{Enc}{}{ECF}\!\left(R, p\right)$ encrypts the confidential data $p$ for the recipients $R$ and returns an \ecf~$\mathcal{E}$.

\sss{Modification of Confidential Data}
Let $p' = \func{modify}(p)$ be the new bit string created by modification of the original confidential data $p$. The replacement of the confidential data within an \ecf~$\mathcal{E}$ is done by these steps: 

\begin{enumerate}[(1)]
    \setlength{\itemsep}{0.25em}
    \item $(R, p) = \f{Dec}{}{ECF}\!\left(\sk{A}{Ed}, \mathcal{E}\right)$
    \item $p' = \func{modify}(p)$
    \item $\mathcal{E}' \leftarrow \f{Enc}{}{ECF}(R, p')$
\end{enumerate}

\sss{Addition of a New Recipient}
Recipient Alice wants to add a new recipient Bob to an existing \ecf. Bob's public Ed25519 key is denoted by $\pk{B}{Ed}$, the bit string of his name by $\f{name}{B}{}$. The signature over Bob's name is denoted by $s = \f{signature}{}{Ed}\!\left(\sk{B}{Ed}, \f{name}{B}{}\right)$. Alice performs the following steps to add Bob to the recipient list: 

\begin{enumerate}[(1)]
    \setlength{\itemsep}{0.2em}
    \item Alice verifies the Signature $s$:\\$\f{verify}{}{Ed}\!\left(s, \pk{B}{Ed}\right) \overset{?}{=} \mathit{Valid}$.
    \item If the signature is invalid, abort the operation, if the signature is valid, proceed.
    \item $\var{r}{B}{} = \left(\pk{B}{Ed}, \f{name}{B}{}, s\right)$
    \item $(R, p) = \f{Dec}{}{ECF}\!\left(\sk{A}{Ed}, \mathcal{E}\right)$
    \item Alice checks whether Bob is already in the recipient list:\\$R \cap \{\f{r}{B}{}\} \overset{?}{=} \emptyset$ {\small(Compare Ed25519 public keys)}.
    \item If Bob is already a recipient, abort the operation, if Bob is not a recipient, proceed.
    \item Optionally, Alice can check if $\f{name}{B}{}$ already exists in one $r_i$ and abort the operation if necessary.
    \item $R' = R \cup \left\{\var{r}{B}{}\right\} = \left\{r_1, r_2, \dots, r_n, \var{r}{B}{}\right\}$
    \item $\mathcal{E}' \leftarrow \f{Enc}{}{ECF}(R', p)$
    \vspace*{-3mm}
\end{enumerate}

\sss{Removal of a Recipient}
Recipient Alice wants to remove a recipient Bob from an existing \ecf. Bob's public Ed25519 key $\pk{B}{Ed}$ and/or the bit string of his name $\f{name}{B}{}$ must be known. If only his name is known, it must be unique within the \ecf~$\mathcal{E}$. Alice performs the following steps to remove Bob from the recipient list:

\begin{enumerate}[(1)]
    \setlength{\itemsep}{0.25em}
    \item $(R, p) = \f{Dec}{}{ECF}\!\left(\sk{A}{Ed}, \mathcal{E}\right)$
    \item Alice searches for $\var{r}{B}{}$ in $R$ based on his public key $\pk{B}{Ed}$ or his name $\f{name}{B}{}$.
    \item If $\var{r}{B}{}$ does not exist (Bob is not a recipient of $\mathcal{E}$), abort the operation, if $\var{r}{B}{}$ exists proceed.
    \item $R' = R \setminus \left\{\var{r}{B}{}\right\}$
    \item $\mathcal{E}' \leftarrow \f{Enc}{}{ECF}(R', p)$
\end{enumerate}

When removing recipients, one does not require any private keys during the encryption process. This implies that recipients of an \ecf\ can remove themselves. It is therefore the task of the implementation to warn the user or abort the operation if the user attempts to do this. Additionally, the implementation should also realize additional security functions if, for example, only the creator of the confidential data stored in an \ecf\ is allowed to add or remove recipients. Finally, it must be noted that the restriction explained in \Cref{sec:relatedwork} is still true: Former recipients are still able to access old versions of an \ecf\ when using a version control system.

\section{Implementation Details}\label{sec:implementation}

A \ac{PoC} was implemented using \cs\ and the \textit{.NET 6.0} runtime. All cryptographic primitives were provided by the portable library \textit{Sodium}~\autocite{Sodium2022}, which is a fork of the \textit{NaCl}~\autocite{Bernstein2016} library. To use \textit{Sodium} with \textit{.NET} a wrapper is needed. For this \ac{PoC} the wrapper library \textit{NSec}~\autocite{NSec2022} was used.

\subsection{Implementation of \ecf\ Functionality}\label{sec:implementation:ecf}

The implementation in \cs\ was subdivided into two projects: \texttt{\small ECF.Core} and \texttt{\small ECF.CLI}. The \texttt{\small ECF.Core} project contains all functionality of the \ecf\ and helps with managing private keys (see \Cref{sec:implementation:privatekey}). \texttt{\small ECF.Core} is a library and can be included into other projects (according to requirements in \Cref{sec:requirements}). It is used by the \texttt{\small ECF.CLI} project which provides a command line interface to the \ecf\ functionality.

The \texttt{\small Create(CipherSuite, ContentType)} function in class \texttt{\small EncryptedContainer} implements the \ecf\ creation process using the given cipher suite and content type. For this \ac{PoC} the aforementioned cipher suite (see \Cref{sec:struct:general}) is implemented as well as a single content type: \textit{BLOB}. Because of the \ac{GCM} mode of operation, the execution platform must support the instruction set extension \textit{AES-NI}. As a rule, this can only lead to problems when using very old processors or virtual machines.

An object of type \texttt{\small EncryptedContainer} can be encrypted using the function \texttt{\small Write(Stream)}. The output is written into the parameter \texttt{\small Stream}. Analogously, one can obtain an unencrypted object of this type using the class function \texttt{\small Load(Stream, ECFKey)}. It is necessary to provide a private Ed25519 key of a recipient to that function. Per default all name's signatures are verified. This can be disabled to achieve better runtime performance during decryption. The property \texttt{\small ContentStream} of an \texttt{\small EncryptedContainer} object provides read and write access to the confidential data.

To protect the private key and the confidential data the implementation uses protected memory spaces, if possible. The library \textit{Sodium} provides suitable functions for this~\autocite{Sodium2022secm}, which in turn are used by \textit{NSec}. Furthermore, heap allocations are replaced by stack allocations wherever possible using the \cs\ keyword \texttt{\small stackalloc}~\autocite{MicrosoftDocs2022}. Alternatively, when using memory that cannot be protected via \textit{Sodium} or stack allocation, the implementation pins these memory regions in memory to prevent the Garbage Collector from arbitrarily copying them. Furthermore, used memory regions are actively deleted before they are freed.

\subsection{Private Key Management}\label{sec:implementation:privatekey}

Using \acp{ECF} requires private keys that should never be stored unencrypted. Therefore, the \texttt{\small ECF.Core} project uses \ac{AES}-256 (\ac{GCM}) to encrypt the private keys. The encryption key is derived from a user-provided password using Argon2id~\autocite{Biryukov2017}\autocite{Sodium2022pwh}. Argon2id aims to enforce costly calculations that cannot be parallelized or otherwise shortened (in time) by an attacker. The algorithm can be configured arbitrarily in order to keep the required computing time variable. For the \ac{PoC} implementation the following settings were chosen:

\begin{itemize}
    \item Degree of parallelism: \tabto{3.5cm} 1 (Limit by \textit{Sodium})
    \item Memory requirements:   \tabto{3.5cm} \qty{2}{\gibi\byte}
    \item Number of iterations:  \tabto{3.5cm} 5
\end{itemize}

This results in an approximate run time of \qty{5}{\second} on an \textit{Intel Core i5-6600K} with newer processors being presumably faster.

The private key is needed when decrypting an \ecf\ and when creating an \ecf. For the latter it is necessary to add oneself as a recipient to that \ecf, which includes signing the name. Therefore, \texttt{\small ECF.CLI} always prompts the user's password to load the encrypted private key.

\section{Conclusion and Future Work}\label{sec:conclusion}
In this paper we introduced \acl{ECF}, a hybrid-encrypted multi-recipient file structure aimed to store confidential data and share it with a customizable set of recipients. Full examples of basic and advanced operations recipients can perform on an \ecf\ were presented in this paper. Although we were using a single cipher suite as described in \Cref{sec:struct:general}, the file format supports multiple cipher suites which can be implemented analogously. The \ac{PoC} implementation demonstrates this by implementing both SHA-512 and SHA-256 as cryptographic hash functions resulting in two different cipher suites.

The full code of the \ac{PoC} implementation and unit tests for that code are available at:
\begin{center}
    \url{https://github.com/Hirnmoder/ECF}
\end{center}
For the future, we plan to add additional cipher suites to \ecf. Additional functionalities are also possible depending on the feedback we get from the community.

\renewcommand*{\bibfont}{\footnotesize}
\setlength{\labelnumberwidth}{0.45cm}
\printbibliography[notcategory=selfref]
\vspace{\fill} 

\refstepcounter{section}\makeatletter\renewcommand{\p@subsection}{Appendix~}\renewcommand{\thesubsection}{\Alph{subsection}}\makeatother
\section*{Appendix}\label{sec:appendix}

\subsection{Generating $m-n$ Obfuscation Blocks}\label{sec:appendix:obfuscation}

In \Cref{sec:struct:encryption} the generation process for the $n$ public recipient-specific blocks was described. The remaining $m-n$ blocks serve as obfuscation blocks to hide the true number of recipients to externals. These obfuscation blocks should not be random bit strings because there is a possibility that the outputs of the used algorithms are subject to statistical effects. This would allow an external party to distinguish between real blocks and obfuscation blocks and therefore determine $n$.

To avoid this, we suggest that the $m-n$ obfuscation blocks are constructed using randomly generated Ed25519 and X25519 key pairs. The function $\f{Gen}{}{Ed}$ denotes the creation of an Ed25519 key pair and the function $\f{Convert}{}{X}\!\left(\sk{}{Ed}\right)$ converts an Ed25519 private key into an X25519 key pair. The following steps are performed for each obfuscation block:

\begin{enumerate}[(1)]
    \setlength{\itemsep}{0.25em}
    \item Generate a random key pair:\\$\left(\sk{r}{Ed}, \pk{r}{Ed}\right)\leftarrow \f{Gen}{}{Ed},~\left(\sk{r}{X}, \pk{r}{X}\right) = \f{Convert}{}{X}\!\left(\sk{r}{Ed}\right)$.
    \item Compute $\func{identification\_tag} = \func{H}\!\left(\pk{r}{Ed} \Vert \func{Salt}\right)\!\trunc{0}{16}$.
    \item Generate an ephemeral X25519 key pair:\\$\left(\sk{e}{X}, \pk{e}{X}\right) \leftarrow \f{Gen}{}{X}$.
    \item Generate a random \ac{AES}-256 key:\\$\aes{r} \leftarrow \f{Gen}{}{AES}(256)$ or $\aes{r} \leftarrow \func{RandomBytes}(32)$.
    \item Execute the key agreement algorithm with the private ephemeral and the random public X25519 keys:\\$\f{k}{shared}{X} = \func{X25519}\!\left(\sk{e}{X}, \pk{r}{X}\right)$.
    \item Compute $\aes{pre2} = \func{H}\!\left(\f{k}{shared}{X} \Vert \pk{r}{X} \Vert \pk{e}{X}\right)\!\trunc{0}{32}$.\\Shortening the hash value to $32$ Bytes is necessary because of the used symmetric encryption algorithm \ac{AES}-256.
    \item Compute $\aes{pre1} = \aes{r} \oplus \aes{pre2}$.
\end{enumerate}

Provided that the used cryptographic hash function generates truly random looking bit strings, on can simplify the generation process to increase runtime performance. The assumption of true random looking bit strings is justified with the input lengths used in \Cref{sec:struct:decryption,,sec:struct:encryption}, see~\autocite{CSE2020}.

\begin{enumerate}[(1)]
    \setlength{\itemsep}{0.25em}
    \item Generate an ephemeral X25519 key pair:\\$\left(\sk{e}{X}, \pk{e}{X}\right) \leftarrow \f{Gen}{}{X}$.
    \item Generate $\func{identification\_tag} \leftarrow \func{RandomBytes}(16)$.
    \item Generate $\aes{pre1} \leftarrow \func{RandomBytes}(32)$.
\end{enumerate}

The shortened generation process is used in the \ac{PoC} implementation. The number $m$ is randomly chosen in dependence on $n$, such that $\max \lbrace8, 2n\rbrace \geq m \geq n$.

\end{document}